\documentclass[12pt,preprint]{aastex}

\shorttitle{Tidally heated moons}
\shortauthors{Scharf}

\begin{document}


\title{The potential for tidally heated icy and temperate moons around exoplanets}


\author{Caleb A. Scharf}
\affil{Columbia Astrobiology Center,
Columbia Astrophysics Laboratory, Columbia University, MC 5247, 550 West 120th Street, New York, NY10027, USA}
\email{caleb@astro.columbia.edu}



\begin{abstract}

Moons of giant planets may represent an alternative to the classical picture of habitable worlds. They may  exist within the circumstellar habitable zone of a parent star, and through tidal energy dissipation they may also offer alternative habitable zones, where stellar insolation plays a secondary, or complementary, role.

We investigate the potential extent of stable satellite
orbits around a set  of 74 known extrasolar giant planets located
beyond 0.6 AU from their parent stars - where moons should be long-lived
with respect to removal by stellar tides.
 For this sample, the typical stable satellite orbital radii
span a band  some $\sim 0.02$ AU in width,
compared to the $\sim 0.12-0.15$ AU bands for the  Jovian and
Saturnian systems. Approximately 60\% of these giant planets can 
sustain satellites or moons in bands up to $\sim 0.04$ AU
in width. For comparison, the Galiean satellites extend to $\sim 0.013$ AU.
 We discuss how the actual number and characteristics of satellites will
depend strongly on the formation pathways.
 
 We investigate the stellar insolation that moons would experience for these
exoplanet systems, and the implications for sublimation loss of volatiles. 
We find that between 15 and 27\% of {\em all}
known exoplanets may be capable of harboring small, icy, moons. In addition, some 22-28\% of all
known exoplanets could harbor moons within a ``sublimation zone'', with insolation temperatures between 273 K and 170 K.
A simplified energy balance model is applied to the
situation of temperate moons, maintained by a combination of stellar insolation
and tidal heat flow. We demonstrate that large moons ($>0.1 $M$_{\oplus}$), at orbital radii commensurate with those of the Galilean satellites, could maintain temperate, or habitable, surface conditions during episodes of
tidal heat dissipation of the order 1-100 times that currently seen on Io.

\end{abstract}




\section{Introduction}

Based on the example of our own solar system, giant planets appear to
be excellent candidates for playing host to significant satellite and
moon systems. Of particular interest is the possibility that moons may
share characteristics with terrestrial type worlds, both in terms of
composition and size, and in terms of thermodynamic conditions
favorable to life. If a host giant planet orbits within the
so-called circumstellar habitable zone then moons of mass
greater than approximately $0.1$M$_{\oplus}$ may be able to retain a
long-lived atmosphere, and sustain temperate surface conditions \citep{williams97}.

Moons may also represent an entirely new class of habitable
environment owing to the potential for them to form beyond the
planetary snow-line in a system and hence accumulate a significant icy
mantle, and the potential for them to enter into orbital resonance conditions favorable for
driving significant tidal heating and sustainable sub-surface liquid
oceans (e.g. Reynolds, McKay \& Kasting 1987).  The latter is a direct 
consequence of the dynamically dense nature of moon systems.
The Galilean moon Europa has served as
a prototype for such objects, and it appears possible that a liquid
water ocean, perhaps as deep as 100 km  (e.g. Melosh et al 2003)
currently exists beneath an outer icy crust.  This state is consistent
with vigorous tidal heating by Jupiter due to the maintenance of an
eccentric orbit ($e\simeq 0.01$) by the mean-motion $4:2:1$ Laplacian resonance
between Io, Europa, and Ganymede. Other moons in our solar system, which would otherwise
be inert, also show evidence for dynamically driven heating. For example,
the recent detection of water ``geysers'' on Enceladus in the Saturnian system (e.g. \citet{porco06}) points
towards a remarkably active geology, even on such a small moon.
In the case of Enceladus, although it is known to be in a 1:2 mean-motion
orbital resonance with the moon Dione (with a resultant orbital
eccentricity of $0.0045$), it appears unlikely that the
resultant tidal heating by Saturn is sufficient to maintain a
sub-surface ocean, although the tides may contribute to localized heating. Proximity
to a librational resonance \citep{wisdom04}, together with radiogenic heating and
a variety of possible internal compositions may provide the additional
boost required to explain it's activity \citep{porco06}.

Non-classical habitable
zones such as that proposed for Europa could greatly extend the potential for biota in planetary
systems by essentially decoupling from the stellar energy output
(e.g.; \citet{reynolds87}) and circumventing the mass requirement for
atmospheric retention. In addition, the classical circumstellar habitable zone
could conceivably be extended when the surface temperature of a 
satellite or moon is maintained by a {\em combination} of stellar insolation
and tidal heating. In the broader context of life within a planetary system such
potential habitats are of enormous interest in seeking both the origins of life
and the capacity for life to survive through unfavorable circumstances. For example,
not only could moons like Europa offer potential ``incubators'' for life, they might - through forward
contamination (e.g. \citet{gladman06}) - offer relatively safe haven for microbial life set adrift
due to cataclysmic events on inner, terrrestrial-type, worlds.

It is not unreasonable to speculate that moon and ring systems will
eventually be detected using more sensitive instrumentation and
current and future observational techniques. Indeed, such detections appear plausible even
with current transit methods ~\citep{sartoretti99}, especially when applied to
future missions such as {\sl KEPLER} or {\sl COROT} which may even be able to
detect the signature of equivalent systems to Jupiter-Europa using eclipse timing (e.g. \citet{doyle03}).
With the eventual construction of 20-30 meter class
telescopes the detection during transit of massive moons and their 
associated atmospheres will also become significantly more feasible \citep{ehrenreich06}.

In this paper we present an initial evaluation of the potential for a subset of currently known
extrasolar giant planets to harbor satellite and moon systems. We also
investigate the likely impact of stellar insolation on icy moons around these exoplanets
by considering the 
orbital incursions within various zones of
stellar insolation, the time-averaged stellar insolation, and the sublimation rates
of water-ice from a moon surface.
 We further consider the potential combination
of stellar insolation and tidal heating of moons in creating habitable
environments, such as temperate moons, and examine the related 
constraints on system parameters.

\section{Satellite orbital terrain}

 Previous works have
investigated the longevity of satellite systems due to tides and
migrations, and in  particular Earth-mass moons around giant planets
on close orbits to the parent star ~\citep{ward73,barnes02}. Barnes \& O'Brien (2002)
note that for parent stars with $M_*>0.15 M_{\odot}$, Earth sized
moons of Jovian planets can remain in stable orbits for some 5 Gyr;
they also point out that beyond approximately $\sim 0.6$ AU there are
essentially no meaningful dynamical constraints on moon masses and survival times. Moons
around such planets could therefore be both massive and long-lived.

In this present work we therefore restrict our investigation to the potential orbital ranges of stable satellite
systems around known planets whose semimajor axes are $a_p>0.6$ AU. 

A simple
consideration of the Hill Sphere radius; $R_H=a_p(M_p/3M_*)^{1/3}$
~\citep{burns86} from the restricted 3-body problem, leads to a broad constraint on the {\em outer}
physical extent of any  system in terms of the satellite
semi-major axis $a_s$. The critical semimajor axis - the location of
the outermost satellite orbit that remains bound to the planet has been 
estimated to be a fraction between $\frac{1}{3}$ and $\frac{1}{2}$ of $R_H$.
We use the numerical result of ~\citet{holman99} for prograde satellites
which suggests a critical semi-major axis of $0.36 R_H$.
For the majority of the exoplanet systems we will consider below, the
planets also have significant orbital eccentricity. We therefore modify our
Hill radius estimate by conservatively considering the {\em perihelion}
star-planet separation (i.e. $(1-e)a_p$) in calculating the allowed satellite
orbits.
Similarly, a consideration of the Roche limit: $d_R=R_s(2M_p/M_s)^{1/3}$, where 
$R_s$, $M_s$ are the satellite orbtial radius and mass respectively, 
 provides a constraint on the {\em inner}
radius of a viable satellite system, such that:

\begin{equation}
\left(\frac{3M_p}{2 \pi \rho_s}\right)^{1/3}\;\; < \;\; a_s \;\; < \;\; 0.36(1-e_p)a_p\left( \frac{M_p}{3 M_*} \right)^{1/3}
\end{equation}

where $M_p$ is the planet mass,
and $M_*$ is the parent star's mass.  In other words --- a moon can
only survive if it is far enough from the planet that it is outside
the satellite-planet Roche limit (which we subsequently label $a_s^{inner}$), is not entirely disrupted,
and close enough to the planet that it is within some fraction of the planet-star Hill
radius (which we subsequently label $a_s^{outer}$). These limits are somewhat conservative, since the Roche limit
is strictly true only for liquid or ``rubble-like'' objects, solid
objects can in principle survive closer in to the host planet. It is
interesting to note that, given the observables $M_p$ (typically modulo $sin
i$), $a_p$, $e_p$, and $M_*$ (from stellar modeling), the only variable which
remains to determine the {\em range} of $a_s$ is the satellite density
$\rho_s$.

\section{The planet sample}

The sample used here has been compiled from both the 
on-line catalog from the Geneva Extrasolar planet search programme
\footnote {\tt http://obswww.unige.ch/\~{}udry/planet/planet.html }, and that
of the Extrasolar Planet Encyclopedia \footnote{\tt http://vo.obspm.fr/exoplanetes/encyclo/encycl.html}
both of which are themselves compilations of detections. We have extracted 
planet candidates (including multiple systems) with semi-major orbital
axes $>0.6$ (see above) and measured eccentricities. We have then
compiled the planet orbital data with the parent star masses estimated
from stellar models, cited in the on-line catalog.  The final sample
used here (at the time of writing) consists of 74 planets, 13 of which are known to
be in multiple planet systems
- excluding these objects does not alter our conclusions.

For all planets we use the estimated $Msini$ as a direct surrogate for
true mass, ignoring therefore any systematics which bias the sample
towards low inclination systems, or the scatter introduced by the
(presumed) random distribution of inclinations. Even allowing for a
50\% ($i=30^{\circ}$) underestimate of $M_p$ our conclusions are not
significantly altered.

In Figure 1 we plot the number of systems with a given range of $a_s$;
$a_s^{outer}-a_s^{inner}$, which we refer to as the allowed satellite
orbital radii or band, estimated from Equation 1. We have assumed here
a mean satellite density ($\rho_s$) of 3 g cm$^{-3}$, commensurate
with the upper-mid range of densities seen in the Galilean satellites
(Io - 3.55 g cm$^{-3}$, Europa - 3.04 g cm$^{-3}$, Ganymede - 1.93 g
cm$^{-3}$, Callisto - 1.83 g cm$^{-3}$)

For this sample the most frequently occurring band is $\sim 0.02$ AU,
and $\sim 60$\% of the giant planets have bands less than 0.04 AU in
width.  For comparison we also indicate in Figure 1 the allowed bands
for the Jovian and Saturnian systems: 0.122 and 0.148 AU
respectively. We also note that, for example, the Galilean moons
actually occupy only the inner 10\% of the allowed Jovian orbital terrain and that it is 
the irregular, smaller, satellites which occupy the outermost
stable orbits. The narrowest allowed band in our sample is
approximately $0.0088$ AU (for the planet HD45350b which has
$Msini=0.98$ M$_{J}$, $a_p=1.77$ AU and the highest eccentricity in
our sample: $e=0.78$).   

These results should not be surprising, the exoplanet sample is strongly biased towards planets
with semi-major axes smaller than those of the giant planets in our own solar system. It does raise the issue however 
of what the impact on moon populations is of a reduced total orbital terrain. We discuss this
in \S7 below

\section{Tidal heating, stellar insolation, and boosted temperatures}

We address three questions: First, what the potential is for Europa-analog moons in the current exoplanet sample, i.e. icy moons beyond the water sublimation line, which are tidally heated to levels commensurate with the presence of subsurface liquid oceans. 
Second, what the present stellar insolation is for hypothetical moons in this exoplanet sample, and what the implications are for the surface environments and retention of volatiles. 
Third, since we find that some $\sim 50$\% of the sample could harbor moons warmer than $170$ K, but not in the liquid-water zone we investigate the requirements for boosting temperatures through tidal heating to attain surface equilibrium  temperatures $273<T_{eq}<373$ K. For this latter question we assume that the moons must be massive enough to retain a
significant atmosphere.

\subsection{Potential Tidal Heating}

We here consider in very simple terms the potential tidal heating for moons due to
the simplest type of mean motion  orbital resonance.
For a satellite or moon in a non-zero eccentricity orbit (
maintained by orbital resonances with other moons - c.f. Io, Europa
and Ganymede, or via resonances with the host planet orbit, see \S8) the rate of tidal dissipation ($\dot{E}$), assuming 
Keplerian motion, synchronous rotation, and zero obliquity, may be written in terms of the surface heat flow
($H_T$):

\begin{equation}
H_T=\frac{\dot{E}}{4\pi  R_s^2}=\frac{21}{38} G^{5/2} \frac{\rho_s^2
R_s^5}{\mu Q} e_s^2 \left(\frac{M_p}{a_s^3}\right) ^{5/2}
\end{equation}

where $\mu$ is the satellite elastic rigidity (assumed uniform), $Q$
is the satellite specific dissipation function, $R_s$ the satellite radius and $e_s$ the
satellite orbital eccentricity, which is assumed to be small (see e.g., \citet{peale80}). 
This may be re-written in terms of the
satellite mass $M_s$ as:

\begin{equation}
H_T=\frac{21}{38} \frac{G^{5/2}}{\mu Q} \left(\frac{3}{4
\pi}\right)^{5/3} e_s^2 \frac{\rho_s^{1/3}}{a_s^{15/2}}
M_s^{5/3}M_p^{5/2}
\end{equation}

For a given $H_T$ it is therefore possible to place a constraint on
the maximum satellite orbital radius where such heating might occur,
given the planet mass and the satellite mass, density, orbital
eccentricity and composition. In order to calibrate this relationship
in a large potential parameter space we first choose a specific question, 
namely; what the potential extent of the
zone is around the planet candidates within which an icy, Europa mass
moon ($\simeq 0.008$M$_{\oplus}$), might be heated to levels commensurate to those
estimated for Europa. These levels
are therefore assumed to be reasonable for maintaining and/or generating subsurface liquid
water oceans.

 We assume that such a moon has an orbital eccentricity commensurate
with that of Europa $e\simeq 0.01$, an icy composition and rigidity
$\mu \simeq 4 \times 10^{10}$ dyne cm$^{-2}$ (appropriate to ice at
temperatures near 100K, c.f. ~\citet{peale80}) and $Q\simeq 100$,
similar to that of solid materials on Earth. We note that the rigidity of a solid, silicate,
rock is a few $10^{11}$ dyne cm$^{-2}$. If a rock-like rigidity was assumed then 
the results presented in this section and those following would
not be significantly altered beyond  factors of 1.4-4 due to lower values of
$H_T$ for a given configuration. In no case would our conclusions be significantly altered.
 Given both the very considerable uncertainty in other
factors, and the varying values of $\mu$ used in the literature, we choose the lower, ice
rigidity in what follows.

We further assume a density
$\rho_s=3$ g cm$^{-3}$.  Estimates of the actual tidal dissipation in
Europa are of course subject to significant uncertainty, however 
surface heat flows of between $H_T\simeq 300 $ ergs s$^{-1}$ cm$^{-2}$ and 
$H_T\simeq 50$ ergs s$^{-1}$ cm$^{-2}$ have been estimated in the
literature ~\citep{cassen79,cassen80}. Such flows are broadly
consistent with the known system parameters and with the  hypothesis
of a tidally sustained sub-surface liquid ocean. For comparison, on Io, the surface heat flow
due to tides is estimated to be $\simeq 1500$ ergs s$^{-1}$ cm$^{-2}$.

The orbital radii at which
such a moon would have to be in order to experience tidal heating
greater than, or equal to $50-300$ ergs s$^{-1}$ cm$^{-2}$ is then estimated. The number of systems versus the 
difference between these radii and the innermost stable orbit ($a_s^{inner}$), is plotted in Figure 2. For the outer
exoplanet sample it is clear that the physical orbital range for such a moon
is commensurate with that of the Jovian system, despite the range of host planet masses,
although in some cases can be a factor 2 larger. In all cases the inner allowed orbit ($a_s^{inner}$)
is less than 0.001 AU. Furthermore, in all systems these levels of tidal heating
can be achieved for orbital radii well within the allowed maximum orbits of Figure 1.  For a given surface heat flow
the required satellite orbit $a_s \propto M_s^{2/9}$. Clearly then, even an Earth-mass
moon would reside only a factor $\sim 3$ times further from the host planet (assuming
all other parameters remain fixed) to attain the same levels of heating as assumed in Figure 2. Only a small
fraction of the exoplanet sample used here ($\sim 3$\%) would not be able to retain such a moon
in a stable orbit. In addition, as described in \S5 below, those systems are also 
those within stellar insolation zones likely unsuitable for icy covered moons with subsurface oceans.
Thus, the reduced satellite orbital terrain in this exoplanet sample (compared to that of the Jovian systems)
 does not appear to limit the potential tidal heating required to sustain putative sub-surface oceans on
 icy moons. Nor does it force such moons to be heated to greater levels, which might result in
 the loss of volatiles (c.f. Io).

 However, if the reduced stable satellite terrain results in a lower likelihood of the presence of 
 large, icy moons then it remains an open question as to whether the exoplanets in the sample
are more or less likely to harbor tidally heated moons. We discuss some of these issues in more detail in \S7 below.

\section{Stellar insolation and moon conditions}

Hayashi (1981) introduced the idea of a planetary snow-line in a proto-stellar/proto-planetary
disk as the distance from the system center at which the local disk gas temperature
drops below 170 K, which is the characteristic, zero-pressure, sublimation temperature for water ice.
Icy moons must almost certainly form beyond this snow-line in a
proto-planetary disk system and beyond the local snow-line within the
circumplanetary disk of the forming host planet. The precise location of the planetary snow line in our
own solar system is
still unclear, indeed it has been suggested that it may have been as small
as $\sim 1$ AU for a dusty disk, compared to the often quoted distance of 4 AU \citep{sasselov00}.

However, a giant planet hosting formed or forming moons with significant icy mantles may migrate inwards to
a final orbit which passes within the planetary snow-line, and/or the later
main sequence evolution of the parent star may result in the water sublimation-line
(so-called to distinguish it from the snow-line in a proto-planetary disk) 
eventually intersecting the planet/moon orbit. In this context it
is appropriate to evaluate the present conditions experienced by
potential moon systems around known exoplanets.

Specifically there are several classes of conditions that ``water rich'' exomoons could experience, based
on sublimation temperatures (170 K), and temperatures commensurate with liquid surface water at one atmosphere pressure (273-373 K) which are typically considered representative of ``habitable'' conditions for life. These classes can be defined as:

\begin{itemize}
\item An orbit, or incursions, within the ``vapor line'', where stellar insolation would produce a mean surface temperature in excess of 373 K.

\item An orbit, or incursions, within a stellar insolation zone resulting in mean surface temperatures between $273$ and $373$ K where, for a massive moon retaining a significant atmosphere \citep{williams97}, liquid surface water could exist, or for a less massive moon, volatile loss would be rapid. We label this as the ``liquid water'' zone rather than the circumstellar habitable zone.

\item An orbit, or incursions, within a stellar insolation zone resulting in mean surface temperatures between $170$ and $273$ K, where significant water-ice sublimation will occur, with subsequent volatile loss for lower mass moons unable to retain an atmosphere.

\item An orbit, or incursions, within a stellar insolation zone resulting in mean surface temperature less than $170$ K, where water-ice sublimation is significantly reduced.

\end{itemize}

By limiting our exoplanet sample to systems with $a_p\geq 0.6$ AU we naturally
avoid planets which spend substantial time within the inner
``vapor-line'' (see below). 

In order to make an initial estimate of the orbital radius at which stellar insolation
produces a given surface temperature we follow the classical prescription for estimating
the equilibrium surface temperature of a fast-rotating body, namely:

\begin{equation}
T_{eq}=\left ( \frac{(1-A_B)L_*}{16\pi \epsilon \sigma d^2} \right )^{1/4} \;\;\;,
\end{equation} 

where $A_B$ is the Bond albedo, $L_*$ the parent stellar luminosity,
and $d$ the distance from the parent star. The factor $\epsilon$ is a crude, first-order, correction
in the case where an atmosphere is assumed (for zero-atmosphere $\epsilon =1$). It incorporates
the infrared optical depth, and for a present-day Earth-type atmosphere $\epsilon \simeq 0.62$ (e.g. \citet{mcguffie05}).
The fast-rotating
approximation should be a fair one when applied to moons of giant planets aligned with
the system orbital plane, since while they should have synchronous
spin-orbits (as is the case for the Galilean satellites) the
combination of their orbital period with that of their host planet
will typically result in rapidly changing, and generally uniform insolation across
the moon surfaces. For moon systems at significant inclination to the system
plane there is the potential for much more static stellar
insolation, we have not considered this situation, but note that the
slow-rotating $T_{eq}$ is only a factor of $\sqrt{2}$ larger than that
for a fast-rotating body.

We have also examined the potential eclipse, or shadowing, times of
moons by the host planet, assuming all moon orbits lie in the planetary system plane. 
These range from $\sim 12$\% to $\sim 0.1$\% of the
total moon orbital period, for close-in ($a_s^{inner}$) and outer ($a_s^{outer}$) moons respectively.
In the case of close-in moons (as defined by Equation 1) the actual
shadow time is typically $\sim 20$ minutes, compared to $\sim 5$ hours
for the outermost allowed moons in the sample. We therefore ignore this effect in
considering the first-order, long term, impact of stellar insolation on moon and
satellite systems.

A major source of uncertainty is whether or not an atmosphere is included
in the estimation of surface conditions. Since we are first concerned here with the potential
vacuum sublimation of surface volatiles, and the impact on moon characteristics, 
we assume zero-atmosphere conditions and set $\epsilon=1$.

Stellar luminosities are compiled for our exoplanet sample from the
on-line catalog and sources described in \S 2. Where luminosities are
not directly available we have estimated them using the reported
optical magnitudes and distances. The total range is $0.29 <
\rm{L}_*/\rm{L}_{\odot} < 4.59$ with a mean of $1.66$L$_{\odot}$.  We
then compute the orbital ranges for $T_{eq}<170$ K, $170$ K$<T_{eq}<273$ K, 
$273<T_{eq}< 373$ K, and $T_{eq}>373$ K,  assuming
albedo's of either $A_B=0.68$, commensurate with that of Europa, or $0.3$ commensurate with that
of a mixed surface, such as the Earth. As described above, we have not
 included the effect of an atmosphere on $T_{eq}$. 


The orbital parameters ($a_p$, $e$, $P$) of the sample exoplanets are used to
evaluate the amount of time each planet spends within a given zone, and to
evaluate time averaged fluxes and temperatures. 
 In both cases we 
are effectively assuming no time latency in reaching an equilibrium 
surface temperature as the stellar insolation varies.

The
situation is summarized in Figures 3 \& 4.  In Figure 3 the number of systems spending
a given fraction of their orbital period within the 3 inner zones is plotted for an assumed $A_B=0.68$. For the zone with
$170$ K$<T_{eq}<273$ K,  exoplanets are {\em not} counted if they enter a zone with $T>273$ K  during
any part of their orbit. In Figure 4 the number of systems as a function of time averaged equilibrium temperature
for objects with $A_B=0.3$ and $0.68$  are plotted.

With an assumed moon albedo $A_B=0.68$  then 22  of the 74
sample planets spend time within the zone of $273<T<373$ K.  Of these,  two 
also spend time {\em interior} to this zone, i.e. within the ``vapor
line''. Beyond this, approximately 53\% of the systems never enter the liquid-water zone and remain
with $T_{eq}< 273$ K.

For those planets passing into, or through, the liquid-water zone
there is a significant range in the amount of time actually spent in
this zone, or within the vapor line. Only 1 system spends its entire
orbital period with the liquid-water zone (depending on assumed albedo), and many
spend less than 30\% of their orbit within this zone, with actual times
varying from a few days to $\sim 100$ days. 73\% of systems spend some time in
the outer ``sublimation'' zone ($170< T_{eq}<273$ K), and 63\% of those do not enter the liquid-water zone. Beyond these zones there are only 18 planets (24\% of the sample) which
always remain beyond their system sublimation-lines.

In terms of time-averaged equilibrium temperatures between 5\% and 18\% of systems
(for albedos of 0.68 and 0.3 respectively) attain $273< T_{eq} < 373$ K, and between
51\% and 28\% of  systems remain at or below the sublimation temperature at all
times (Figure 4). The remaining 43\% or 54\% of systems have surface temperatures which place them
in  what might be best termed the sublimation zone ($170$ K$<T_{eq}<273$  K) - a similar number to that
determined from the simple zoning criteria above.

Allowing for atmospheres ($\epsilon < 1$) in Equation 4 would shift all insolation zone radii
a outwards (e.g. by a factor $\sim 1.27$ for $\epsilon=0.62$). However, our primary purpose here is to evaluate the 
impact of insolation on the icy mantle of {\em small} moons, where atmospheres may be hard to
retain. We do note however that for those systems which may remain within
the liquid-water zone at all times, or with an equivalent time-averaged temperature,
then one would expect that if an atmosphere can
be retained by a moon then substantial liquid surface water could potentially
exist \citep{williams97}. 

\subsubsection{Sublimation rates \& volatiles loss}

As described above, in vacuum the sublimation temperature for pure water ice is $\sim 170$ K, and
this is the physical criterion used to define the sublimation-line in a planetary
system. A significant number of the exoplanets in our
chosen sample spend between 10\% and 100\% of their orbital periods within this
sublimation-line, but outside of the liquid-water line. To first order we can evaluate the effect of
this on an icy mantle on associated moons by considering the upward sublimation
rate of water ice at these temperatures. Following \citet{spencer87} we plot in Figure 5
 the rate of surface lowering due to sublimation for water ice as a function 
of temperature. The water vapor pressure over ice is used in this calculation.
This plot assumes {\em only} sublimation loss, and does not
account for re-deposition of material, which at $\sim 170$ K may be at rates approximately equal to those
of sublimation - depending on local 
surface temperature variations (e.g. latudinal variation on Galilean satellites, 
\citet{spencer87}).
What is immediately apparent is that for $T_{eq}>170$ K sublimation loss
rates are extremely quick - with 100km depth of water ice sublimating
in only a few $10^6$ years at 170 K - if there is no re-deposition.

The escape velocity from the surface of a Europa mass moon ($\sim
0.0082$M$_{\oplus}$) is $\sim 2$ km s$^{-1}$, compared to a mean
velocity of a water molecule in a gas at 170 K of $\sim 0.4$
km s$^{-1}$.  Applying the thermal (Jeans) escape methodology
(e.g. \citet{lammer04}) then the typical flux of escaping gas particles at
these temperatures is at least a factor $10^8$ lower than that for gas
in the exosphere of a large moon with 1000 K temps (e.g. similar to the
Martian exosphere). Thus, thermal escape appears unlikely to be a
dominant mechanism for material loss in cold moons beyond the
sublimation line, and even up to the ice-line at 273 K. However, as
\citet{williams97} describe, sputtering by charged particles
trapped within the magnetosphere of the giant planet host is likely to
be a highly efficient loss mechanism for moons - if they do not have
protective magnetic fields. \citet{williams97} further
suggest that moons of mass $<0.1$ M$_{\oplus}$, without magnetic
fields, at {\em any} distance from
the parent star will have a hard time retaining an atmosphere for as
long as a billion years.

Thus, it appears reasonable to speculate that small ($<0.1$M$_{\oplus}$)
icy moons which spend significant time
within the sublimation-line are likely to lose sublimated material and
eventually all surface volatiles over relatively short timescales. The major
caveat to this statement is that {\em if} a small moon with a host planet in such
an orbital configuration has a strong magnetic field it might retain a
cold atmosphere of volatiles, and their dissociated atomic species over
longer timescales. 

For our current sample of exoplanets, and the zones occupied as described in \S5, between
49\% and 72\% of these systems are such that small moons are unlikely to have retained any surface water from  an icy mantle - if they initially had one. Thus, we can re-evaluate our estimate of the potential for small, tidally heated ``Europa-like'' moons to
restrict ourselves to those beyond the sublimation line. As seen in \S 5 this would result in some 28-51\% of our current sample of exoplanets being capable of harboring small, icy, moons with the potential for tidally heated subsurface oceans. In terms of
{\em all} currently known exoplanets (i.e. including those within 0.6 AU of their parent star) this corresponds to 15-27\% of the total population.

\subsection{Tidally boosted temperatures}

An intriguing possibility, raised by \citet{reynolds87}, is that
a {\em combination} of stellar insolation and tidal heating act to raise the
surface temperature of a moon to within the classically habitable range.
The range of stellar insolation seen around the above sample of exoplanets is also
motivation for examining this possibility, since many planets are only just beyond the liquid water zone of their parent
stars.
To examine this we make a very naive assumption that the equilibrium
temperature of an object's surface - where we define the surface as some
{\sl ad hoc} layer of outer material - is that of a pure black body receiving
both an input radiation flux {\em and} an input flux from tidal heating. 
This condition can be written as a form of zero-order energy-balance equation;

\begin{equation}
\epsilon \sigma T^4_{eq}= \frac{(1-A_B)}{4} f_{rad} + H_T\;\;\;,
\end{equation}

where $f_{rad}$ and $H_T$ are the stellar flux and tidal surface
heat flow respectively. An implicit assumption is that the tidal surface heat flow
(which is really just the net rate of tidal energy dissapation divided by the moon
surface area) acts exactly like a radiation field. In other words the energy flux
is assumed to be entirely thermalized by the object's surface.
This is extremely crude, by comparison \citet{reynolds87}
employed a radiative-convective equilibrium code to examine the impact of
tidal heating on a hypothetical object similar to Titan (i.e. with a thick
atmosphere). However, the {\em differential} adjustment to $T_{eq}$ as
$f_{tidal}$ is varied is quite similar (within an order of magnitude) between the more sophisticated model
with an atmosphere and our simplistic model, where a ``greenhouse'' atmosphere is
incorporated via the atmospheric infrared transparency factor $\epsilon$ in Equation 5. We note as well
that neither model allows for non-thermal energy dissipation - e.g. the tidal surface
energy flow could equally power the bulk rearrangement of an object's surface.
We therefore use the above to illustrate in the broadest terms the potential for
tidal boosting.

In Figure 6 we present some of the constraints and
potential environments discussed here in terms of the parameters $M_s$
and $a_p$, the satellite or moon mass, and the distance from the parent
star respectively. Major limits on $M_s$ and $a_p$ are indicated by
shaded regions. The approximate minimum mass for retention of an
atmosphere is indicated at $M_s=0.1$M$_{\oplus}$ (e.g. \citet{williams97}).
 The limiting orbital semi-major axis below which moons will be
tidally stripped on relatively short timescales is also indicated at
$a_p=0.6$ AU (e.g. \citet{barnes02}).  Immediately beyond this
line is the ice-line ($T_{eq}=273$ K) for an assumed $1$L$_{\odot}$
parent star, an albedo $A_B=0.68$, and $\epsilon=0.62$, which occurs at 0.75 AU. The
related vapor and sublimation lines are also plotted. We also plot a
hypothetical upper mass limit based on the minimum mass subnebula
postulated for the Jovian system \citep{canup02}. The subnebula has been
estimated at $\sim 2$\% of the final mass of Jupiter, or $\sim
6$M$_{\oplus}$, and we therefore treat this as a zero order estimate of the maximum possible
mass for a single moon.

 Given the expression for tidal heating in Equation 3,
we can substitute for $a_s$ by either the minimum allowed satellite orbital
radius ($a^{inner}_{s}$) or the outer, $a_{s}^{outer}$. In both cases the
host planet mass $M_p$ cancels out and we can rearrange for $M_s$, namely:

\begin{equation}
M_s^{inner}= \left(\frac{38}{21}\right)^{3/5} \left(\frac{6}{\pi}\right)^{1/2} \frac{(\mu Q)^{3/5}}{G^{3/2}} \left[\frac{H_T}{e_s^2 \rho_s^{17/6}} \right]^{3/5} \;\;\;,
\end{equation}

where we have assumed $\mu Q= 4\times 10^{12}$ dyne cm$^{-2}$ in what follows. Similarly for the outer
allowed satellite orbital radius we obtain:

\begin{equation}
M_s^{outer}=\left(\frac{38}{21}\right)^{3/5} \frac{4\pi}{3}\frac{(\mu Q)^{3/5}}{G^{3/2}} \left[\frac{H_T ((1-e_p)a_p)^{15/2}}{e_s^2 \rho_s^{1/3} (3M_*)^{5/2}} \right]^{3/5} \;\;\;.
\end{equation}

$M_s^{inner}$ and $M_s^{outer}$ are therefore the required moon masses at the inner and outer
allowed orbital radii to generate a given $H_T$, assuming all other parameters are fixed.
It is immediately apparent that $M_s^{inner}$ will be small for
most interesting choices of parameters. Equation 7 can also be scaled for moons at
arbitrary radii within the outer allowed orbital radius. Three sets of curves (dashed and solid lines) are plotted in Figure 6 corresponding to loci of constant $T_{eq}$ (373 K and 273 K respectively) for moons at radii 0.1$a_s^{outer}$, including the effects of stellar insolation according to Equation 5 (and $A_B=0.68$, $\epsilon=0.62$). This orbital radius scaling is chosen to
correspond with the approximate relative location of the Galiean satellites in the Jovian system (Figure 1).
 As described above, we choose $\rho_s=3$ g
cm$^{-3}$, and from left to right increasing values of moon
eccentricity: $e_s=0.001$, 0.01, and 0.1. Zero eccentricity, $e_p=0$, is assumed for the planet. 
The regions between the pairs of curves for a given $e_s$ therefore represent the
temperate (habitable) zone produced by the combination of stellar insolation
and tidal heating - which clearly extends this zone to greater
distances (by as much as a factor $\sim 2$ in planet-star radius for an Earth mass moon with $e_s=0.01$) from the parent star than the classical, stellar insolation, zone alone. 

However, in addition to the major caveat that the ``energy balance''
represented by Equation 5 is extremely approximate, it should be
noted that the actual levels of tidal heating required in these zones
can be as high as $\sim 10^6$ erg s$^{-1}$ cm$^{-2}$ - which is a factor
$\sim 10^3$ larger than the surface heat flow estimated for Io. It is an interesting
question whether or not this would create such an unstable surface environment
that habitability would be compromised. More modest heating levels 
are required to boost temperatures which are already close to temperate.

\section{Tidal boosting and known exoplanet systems}

The outer stable orbit (Equation 1) provides a natural scaling for a moon system architecture. A given hypothetical moon must have a semi-major axis of $\beta a_s^{outer}$ (assuming small
eccentricities), where $\beta \leq 1$ and $\beta a_s^{outer}\geq a_s^{inner}$. We can also write this in terms of tidal heat flow:

\begin{equation}
\beta a_s^{outer}= \left(\frac{21}{38}\right)^{2/15} \left(\frac{3}{4\pi}\right)^{2/9} \frac{G^{1/3}}{(\mu Q)^{2/15}} \left[ \frac{e_s^2 \rho_s^{1/3} M_s^{5/3}M_p^{5/2}}{H_T}\right] ^{2/15} ;\;\;.
\end{equation}

Thus, for a given exoplanet in the sample used here we can estimate the time-averaged $H_T$ required to attain
a given $T_{eq}$ (Equation 5), and for an assumed set of moon properties, such as mass, orbital eccentricity, density, rigidity, dissipation, and atmosphere we can then evaluate the required orbital radius of the moon $\beta a_s^{outer}$.

We have applied this calculation to the subset of known exoplanets considered above. Figure 7 summarizes the results for the example of a $0.1$ M$_{\oplus}$ moon (the minimum mass moon likely capable of retaining a terrestrial-type atmosphere) with an
orbital eccentricity of $e_s=0.01$, density $\rho_s=3$ g cm$^{-3}$, and albedo, rigidity and dissipation commensurate with that 
estimated for Europa. We have assumed an atmosphere with $\epsilon=0.62$. To attain a moon surface temperature $T_{eq}=273 K$ then (as described above) 46 of the exoplanets  would require a moon to have  $H_T>0$ to boost the stellar insolation, for $T_{eq}=373 K$ then 70 of the exoplanets would require a moon to have $H_T>0$. In Figure 7 we plot the distribution of both the absolute orbital semi-major axis required for such moons and the ratio of this orbital axis to the {\em inner} ($a_s^{inner}$) stable orbital semi-major axis. The latter plot confirms that such moons would reside comfortably outside of the
inner Roche limit.

For the range of habitable surface temperatures the required orbital range is $\sim 0.002 - 0.007$ AU, corresponding to
$\sim 4-10$ times the innermost stable orbital radius. Since $\beta a_s^{outer}$ only scales as $M_s^{2/9}$ a larger moon of mass $1$M$_{\oplus}$ would shift these ranges outwards by approximately 67\%.  In either case, the relevant orbital terrain is very similar in absolute terms, to that occupied by the Galilean moons (as explicitly indicated in Figure 7), and does not therefore raise any immediate concerns that this would be an unusual configuration for a giant planet.

The tidal surface heat flow $H_T$ required in Figure 7 ranges from $\sim 10^3  - 10^5$ erg s$^{-1}$ cm$^{-2}$, compared to (for example) the $1.5 \times 10^{3}$ erg s$^{-1}$ cm$^{-2}$ estimated for Io. In Figure 8 the distribution of total energy dissipation rates for the case of $0.1$M$_{\oplus}$, $e_s=0.01$ and $T_{eq}=273 K$ is shown, and ranges from $\sim 10^{21} - 10^{23}$ erg s$^{-1}$. By comparison, for Io the total dissipation rate is $\sim 6 \times 10^{20}$ erg s$^{-1}$. These levels of energy dissipation for the boosted moons therefore raise a number of questions. First, very basic order of magnitude energy estimates suggest that  dissipation at a level of $10^{23}$ erg s$^{-1}$ may not provide enough longevity to be of interest for habitability. The {\em total} energy of a planet-moon system consisting of a Jupiter mass planet and a 0.1M$_{\oplus}$ moon with 0.003 AU orbital semi-major axis is dominated by the rotational energy of the giant planet. In the case of Jupiter this total
energy is at least $10^40$ ergs, modulo uncertainties in internal rotation and composition.
This energy could be dissipated completely in $\sim 3$ billion years at a rate of $10^{23}$ erg s$^{-1}$ - assuming tidal heating was sustained at this level, whereas in reality the moon might well move out of the assumed mean-motion resonance on a shorter timescale. It is also not yet known whether Jupiter's apparently fast rotation rate is actually typical for giant planets - others may spin slower.
For the lower end of dissipation energies the maximum lifetime could be one or two orders of magnitude larger, however the issue remains that the mean-motion resonances may not last that long, and indeed may come and go over a moons lifetime.  Second, as discussed in the previous section, it is unclear exactly how these levels of tidal heating would manifest themselves in terms of surface environment. For a liquid ocean dominated moon, the bulk motions would be considerable. For a moon where tidal heating results in significant tectonic and volcanic activity, the surface environment might be rendered less hospitable as a result.

\section{Satellite formation and orbital terrain}

We discuss briefly the relationship of surviving moon or satellite
systems to the formation and early history of a given planetary system
in general. There is currently no clear consensus on the precise
formation mechanisms for satellites of giant planets - although it is
certainly very different from the mechanism likely responsible for
producing the Earth-Moon system, \citep{peale99}. The standard model
(e.g. see review by \citet{pollack91}) for the
Galilean satellites follows  an analog to the core-accretion model for
giant planets, but invokes a minimum mass {\em sub}nebula - where the
total observed  mass in circumplanetary solid material is added to sufficient gas to
match solar element abundances. However, dynamical  timescales in
satellite systems are much shorter than their planetary analogs and it
is not clear that such a model can succeed in reproducing the known
Galilean satellites and their often icy compositions. Alternatives
include the model of \citet{canup02} which invokes a gas-starved
circumplanetary accretion disk where satellites form late relative to
the growth of the giant planet (specifically Jupiter). In both scenarios the 
migration of satellites within the circumplanetary disk can proceed much
as it does for the planets themselves. 

Based on the above results, while we have shown that the orbital
terrain suitable for significant tidal heating due to planet-moon
interaction in the existing exoplanet sample is physically comparable to
that in the Jovian system, the actual total stable
satellite orbital terrain is typically much smaller by a factor $\sim
5-6$. This raises the question; does this reduced  terrain also reduce
the number of potential satellites/moons and therefore reduce the net
likelihood of tidally heated moons ?

As already noted (Figure 1), the actual occupancy of the stable satellite
orbital terrain in the Jovian system is small for the Galilean satellites, which
span only the inner $\sim 10$\% of the total allowed range. The lesser satellites, which
are a factor 10-100 smaller in radius, do however occupy the terrain out to $a_s^{outer}$.

 However, the  bigger issue is whether an analogous system of large
moons could form around a planet with, for example, a {\em total}
stable orbital terrain some $\sim 16-20$\% of the Jovian range. In
order to answer this question we need to know (amongst other factors)
whether or not a planet has undergone significant migration toward the
parent star, thereby reducing the stable satellite terrain, and how
the timing of such a migration compares to the formation time of a
satellite system. Two basic, and undoubtedly simplistic, scenarios can
be considered which more or less bracket the range of possible
models. In the first, the satellite system of a giant planet forms
essentially contemporaneously with the planet, and in the case of a
migrating planet, during the period of either Type I or II
migration. In the second, the satellite system forms ``late'' in a
largely gas-free environment due to N-body accretion
(e.g. \citet{canup00}), and at a time when the host planet has ceased
migration. In the first scenario, although the planet-star Hill sphere
radius shrinks as the planet migrates inwards the circumplanetary disk
continues to receive inflowing material (e.g. \citet{lubow99}) - thus
forming satellites can continue to accrete material even if the stable
orbital terrain is shrinking. In the second scenario the formation of
satellites is potentially constrained by the amount of material
retained according to the Hill sphere criterion (e.g. Equation 1). If
the first scenario occurs then we can argue that a reduced final
stable orbital terrain compared to, for example, the Jovian system,
does {\em not} necessarily reduce the likelihood of the formation of
an equivalent number of large moons. In this case then it will be
correct to state that there is at least an equal potential for tidally
heated moon systems in the current exoplanet sample compared to the
Jovian system - as summarized by Figure 2.  This is also clearly dependent on
the actual sub-nebula mass, which we have assumed scales with the
final host planet mass. However, in the second
scenario it may well be the case that fewer moons can be formed in
which case certain exoplanets may represent a more barren terrain for
significant moons, heated or otherwise. It should be noted that a third scenario
exists, which is that moons are captured bodies which did not participate in the
evolution of the sub-nebula. 

The issue of moon composition, in particular the presence of an icy mantle,
has been briefly described in \S5.0.1. In addition to issues of sublimation rates,
a full treatment would necessarily include discussion of the chemical/elemental
structure of the proto-planetary nebula, and sub-nebula. The distribution of
water in proto-planetary systems is clearly relevant in this case.


\section{Summary and Discussion}

We present an initial investigation of the potential for, and possible scenarios of, moons around giant exoplanets
which are subject to tidal heating and which may therefore provide a distinct class of habitable environment beyond that of the classical ``habitable zone''.
For the majority of $a_p>0.6$ AU planetary systems discovered thus far,
we make a clear prediction that any  associated, long-lived, satellite
systems will occupy a significantly narrower range of  orbital
semi-major axes than the examples of the Jovian or Saturnian
systems. Despite this, the potential for tidal heating to levels sufficient to sustain
sub-surface liquid water oceans does not appear to be impacted, even for Earth-sized moons - {\em if}
the formation pathways to large moons are not effected by the narrower stable orbital terrain.

We have also explored the zonal and time-averaged stellar insolation that  hypothetical moon systems around known exoplanets might experience. In many cases it seems likely that if  ice shrouded moons had at some time
existed around these planets then sublimation processes will have likely removed this material. For small moons the volatiles will have been lost, for moons more massive than some $0.1$ M$_{\oplus}$ \citep{williams97} an atmosphere may have been retained. Between 28 and 51\% of the exoplanet systems studied here are expected to be capable of harboring
small moons with intact icy mantles. This translates to between 15 and 27\% of the {\em total} population of known exoplanets at this time.

We consider the possibility of a combination of stellar insolation and tidal ``boosting'' raising the surface temperature of a moon into the temperate range ($273 < T_{eq} <373$ K). A very general set of constraints is presented; demonstrating how this can lead to an extended temperate, or habitable, zone of as much as a factor $\sim 2$ greater distance from the parent star for massive ($\sim 1 $M$_{\oplus}$) moons. We also find that the relevant orbital terrain around the known exoplanets for tidally boosted, temperate, massive moons is essentially the same as that of the Galilean satellites in the Jovian system. However, the required tidal heating energy budgets range from the level seen in Io to as much as 100 times greater. At this upper extreme it is not clear whether such dissipation could be either long-lived (more than a few 100 million years) or compatible with a habitable surface environment.

We have ignored a multitude of possible additional factors in making
our estimates - for example, there are likely significant resonance
conditions which may occur for moons orbiting planets which are themselves in
eccentric orbits about the parent star, and there are also likely
influences from other, as yet undetected, planets in the system. We have also not
considered resonances beyond simple orbital mean-motions - for example,
 spin-orbit librational resonances such as that likely in the case of Enceladus \citep{wisdom04}.
Such effects are
interesting and should be explored further. We have also not included any
estimate of the {\em likelihood} that moons will enter into resonance conditions that
will drive orbital eccentricity and hence moon-planet tidal dissipation - or of the dynamical timescales
of such situations. This
will require a more extensive study which should include modeling of the
origin of a satellite or moon system and the potential orbital migration of major
satellites within such a system. Such a study should also include an evaluation of 
 any variation in the composition and structural
properties of the satellites
resulting from different circumstellar/circumplanetary formation distances.

However, taken at face value, one of the implications of the
above observations is that the potential for sub-surface oceans in
icy moons, or even ocean moons (for those systems which lie close
enough to the parent star for significant surface temperatures to
develop, boosted by tidal dissipation) suggests that broad questions of habitability may need to be
re-visited to include such environments as significant potential
biospheres. As the parameter space for exoplanet detection expands to
include lower-mass, larger-orbital-radii systems, and if future missions and instruments
begin to detect the presence of moons around giant planets, we will  be
able to extend the evaluation of such potential habitats.

\acknowledgments The author acknowledges the funding support of the Columbia Astrobiology Center through
Columbia University's Initiatives in Science and Engineering, and the
support of the Columbia Astrophysics Laboratory. This work is also
directly supported by a NASA Astrobiology: Exobiology and Evolutionary Biology; and Planetary Protection Research grant, \# NNG05GO79G. The referee, Jason Barnes, is thanked for comments that have drastically improved the manuscript.
Kristen Menou and David Helfand are also
thanked for useful discussions and suggestions.

\clearpage

\begin{figure}
\plotone{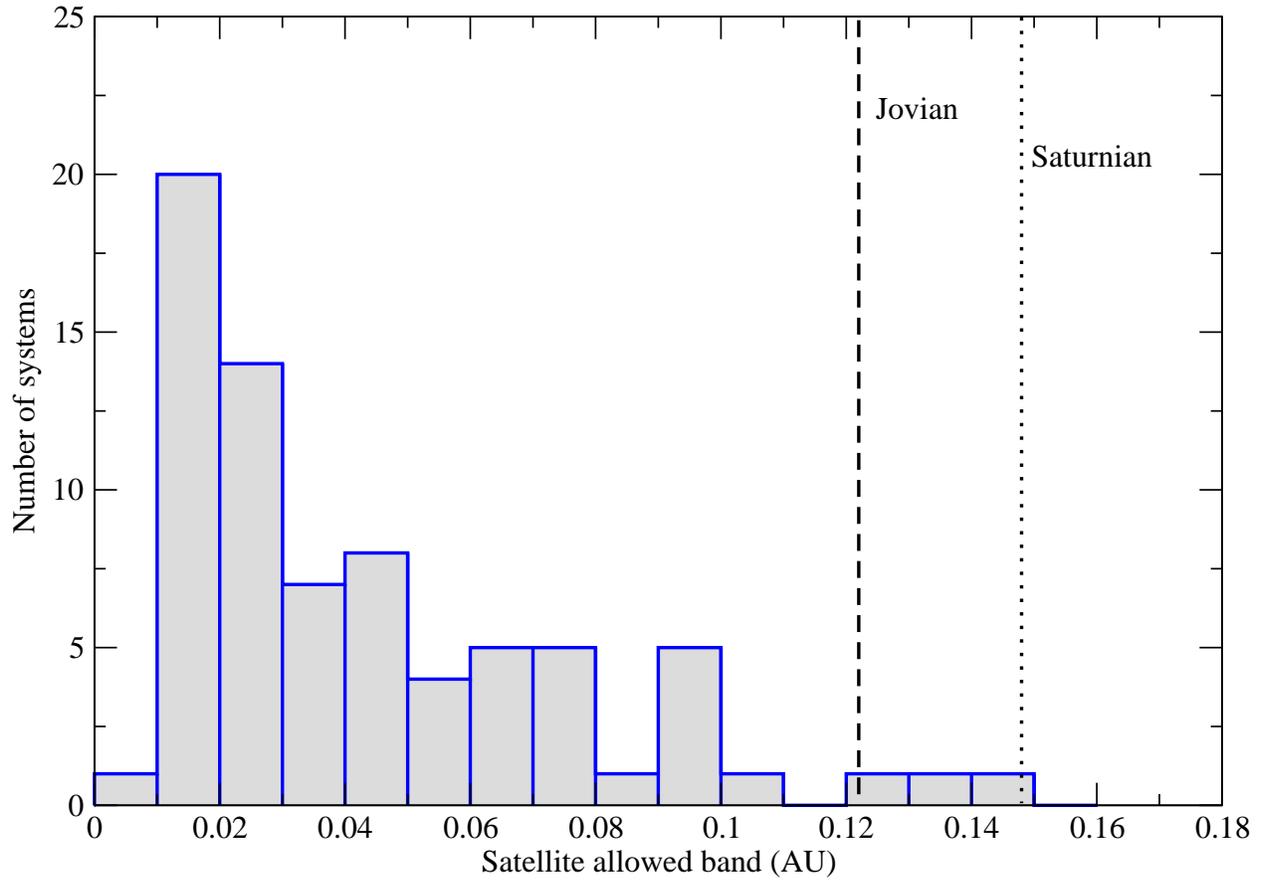}
\caption{The number of systems as a function of stable satellite orbital
  radii band size. Bins are 0.05 AU in width. The total sample of 74
  systems with semi-major axes $>0.6$ AU is included. Vertical dashed, dotted lines represent the
  Jovian and Saturnian band widths respectively.
\label{fig1}}
\end{figure}

\begin{figure}
\plotone{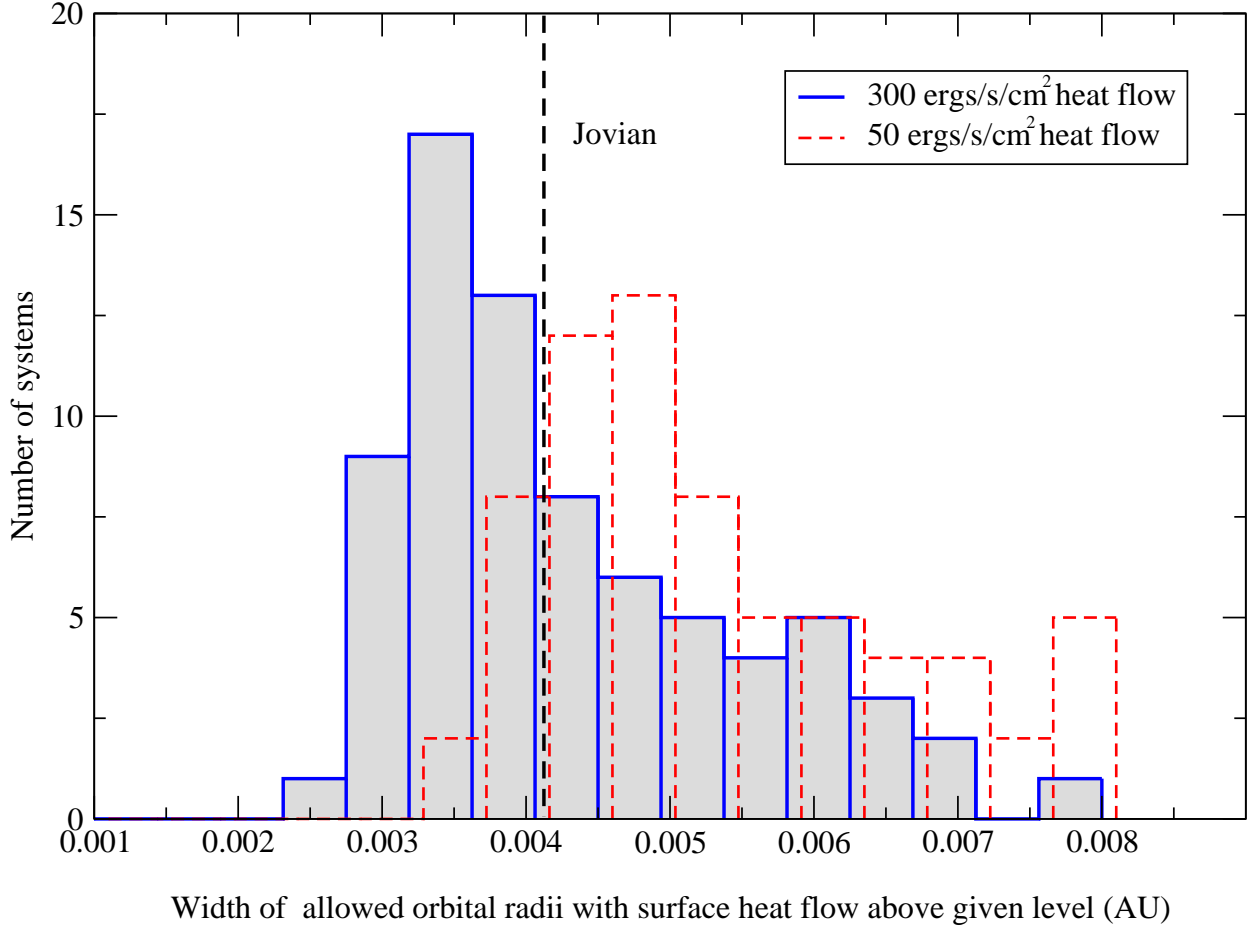}
\caption{The number of systems as a function of the width of the
allowed orbital radii (in AU) within which surface heat flow due to tidal
heating of a moon with a mass ($0.08$M$_{\oplus}$), composition and  orbital
eccentricity ($e_s=0.01$) equal to that of Europa, is greater than or equal to the range of
values inferred for Europa: $300$ ergs s$^{-1}$ cm$^{-2}$ (filled
bars, blue outline) and  $50$ ergs s$^{-1}$ cm$^{-2}$ (dashed red
outline). Vertical dashed line corresponds to the orbital radius of
Europa minus the inner Roche limit radius.
\label{fig2}}
\end{figure}

\begin{figure}
\plotone{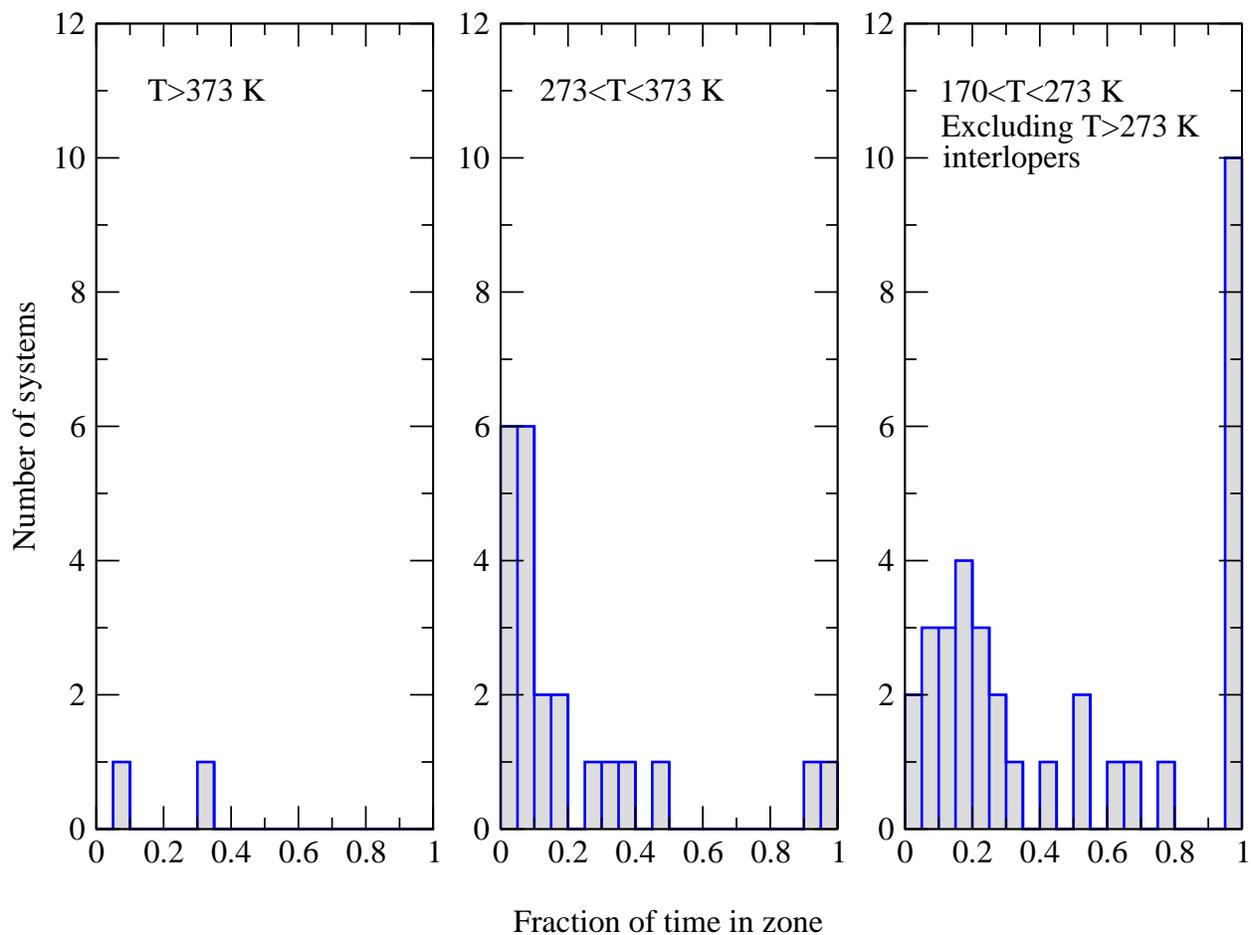}
\caption{The number of exoplanet systems as a function of the fraction of total orbital period spent within a particular stellar insolation zone. The three innermost insolation zones are shown in panels (a) $T_{eq}>373$ K, (b) $273<T_{eq}<373$ K, (c) $170<T_{eq}<273$ K. In panel (c) planets are not counted which enter the $273<T_{eq}<373$ K zone at any point in their orbit in order to assess the number which remain within the sublimation zone, or beyond, at all times. }
\label{fig3}
\end{figure}

\begin{figure}
\plotone{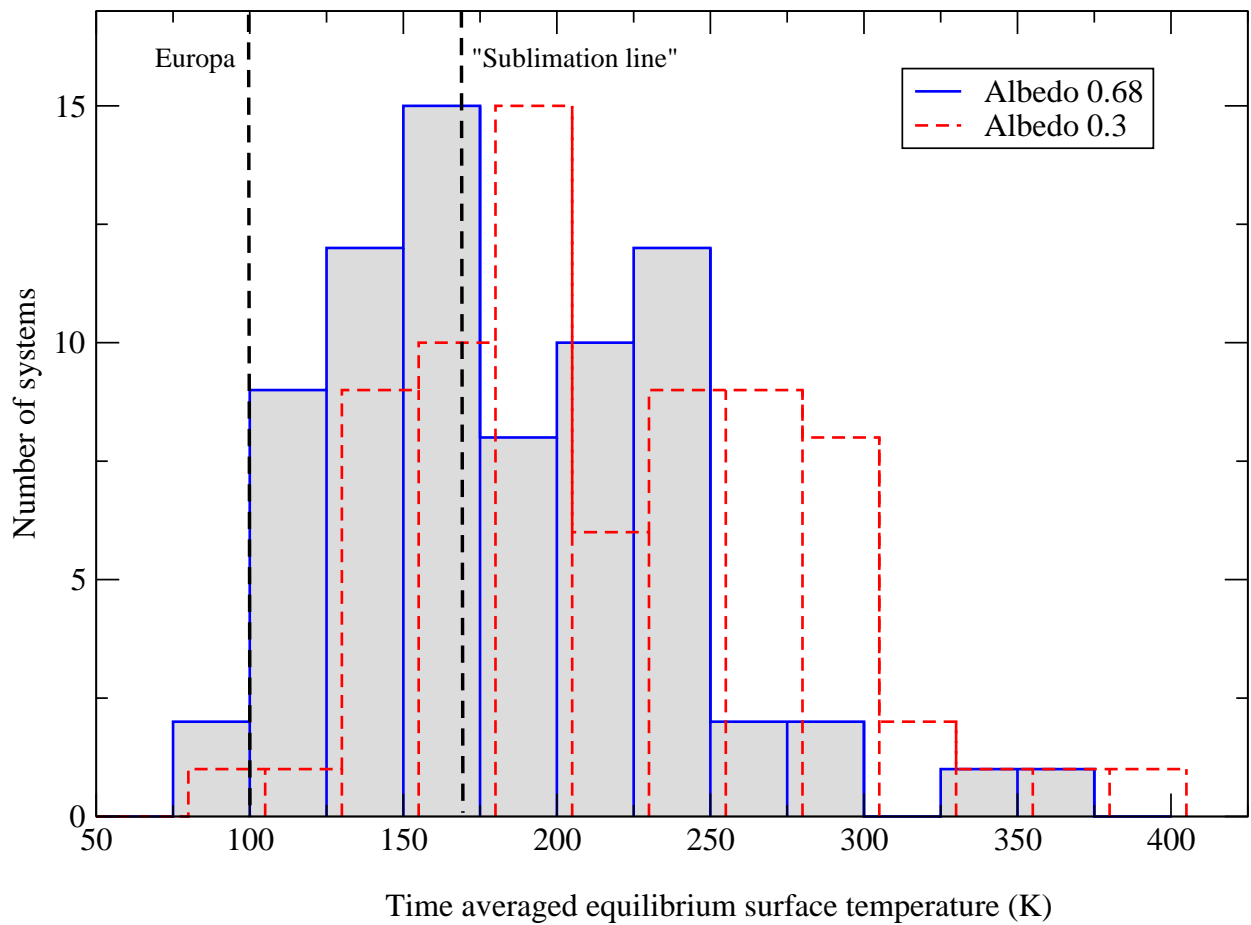}
\caption{The distribution of time-averaged surface temperatures due to stellar insolation for atmosphere-free moons around the sample of 74 exoplanets used here. Two albedo's are used; $A_{B}=0.68$ commensurate with an icy moon such as Europa, and $A_{B}=0.3$ commensurate with a mixed surface, such as that of a cloud free Earth. The water sublimation line is plotted at 170 K (dashed vertical line), as well as the mean temperature of Europa at 100 K (dashed vertical line).}
\end{figure}

\begin{figure}
\plotone{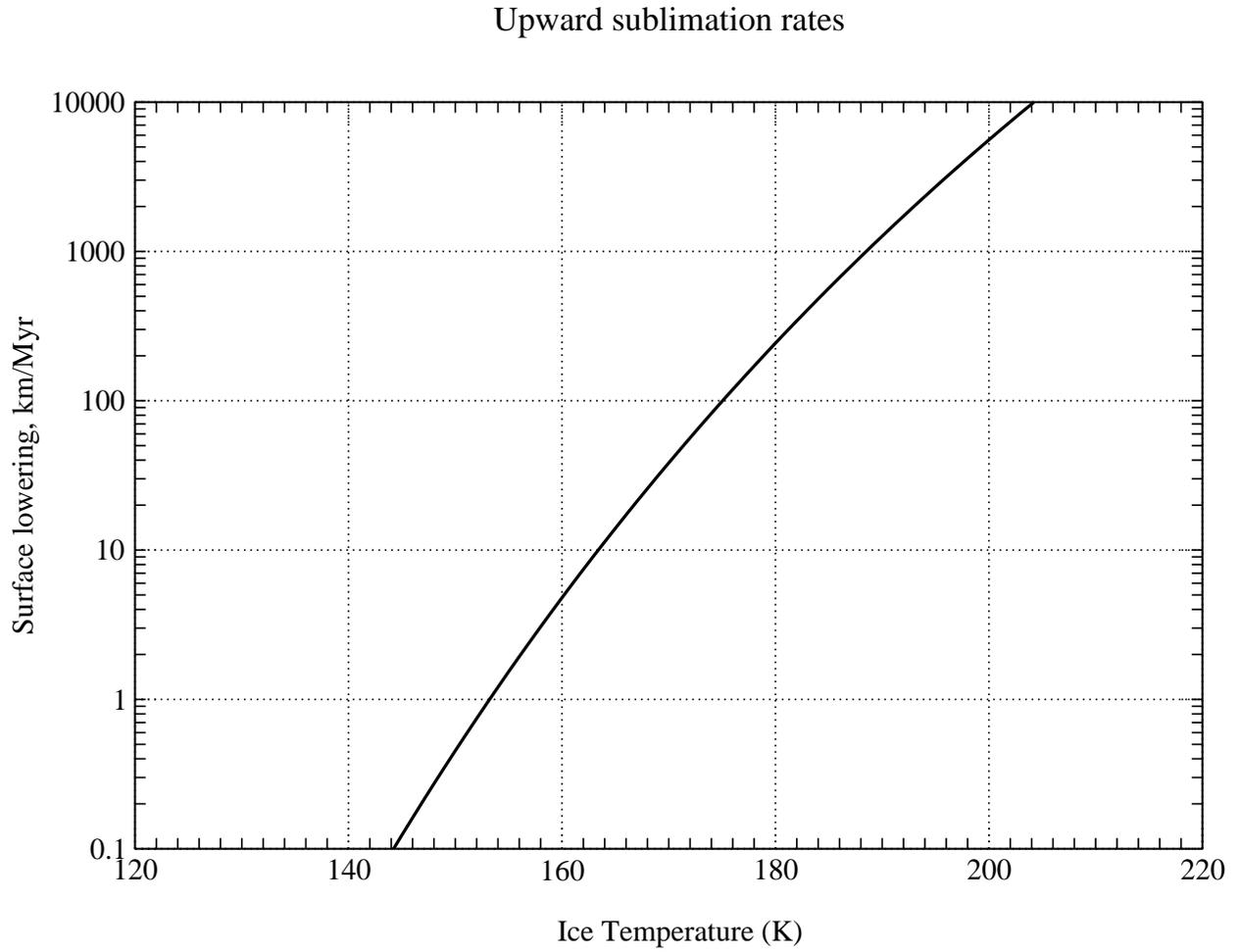}
\caption{The upward sublimation rate of water ice in vacuum as a function of temperature expressed
in terms of the lowering of a surface in kilometers per $10^6$ years. The rate is estimated using the water vapor pressure over ice, following Spencer (1987).}
\end{figure}

\begin{figure}
\plotone{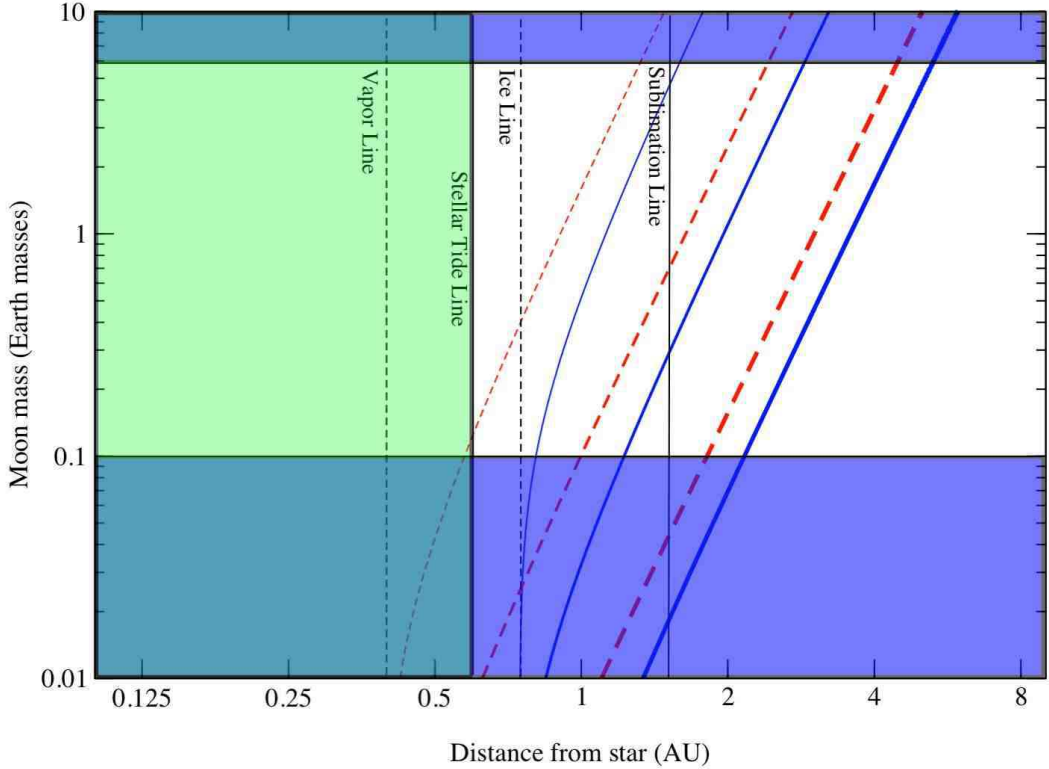}
\caption{Constraints on possible moon masses and environments as a function of distance from parent star. Shaded area below 0.1 M$_{\oplus}$ indicates region of moons likely incapable of retaining an atmosphere. Shaded area to left of
solid vertical line labeled ``Stellar Tide Line'' at 0.6 AU indicates planet-star separations where long-lived moons are unlikely. Upper shaded area above 6M$_{\oplus}$ indicates possible upper limit to moon mass based on Jovian minimum-mass sub-nebula. Vertical lines labeled ``Vapor Line'', ``Ice Line'', and ``Sublimation Line'', correspond to 
locations of equilibrium temperature zones of 373 K, 273 K, and 170 K for an assumed albedo of $A_B=0.68$ and
atmosphere with $\epsilon=0.62$. Dashed (red) and solid (blue) curves correspond to loci of temperatures 373 K and 273 K attained via tidal boosting assuming (left to right) $e=0.001$, $0.01$ and $0.1$ with $A_B=0.68$, $\epsilon=0.62$.
}
\end{figure}

\begin{figure}
\plotone{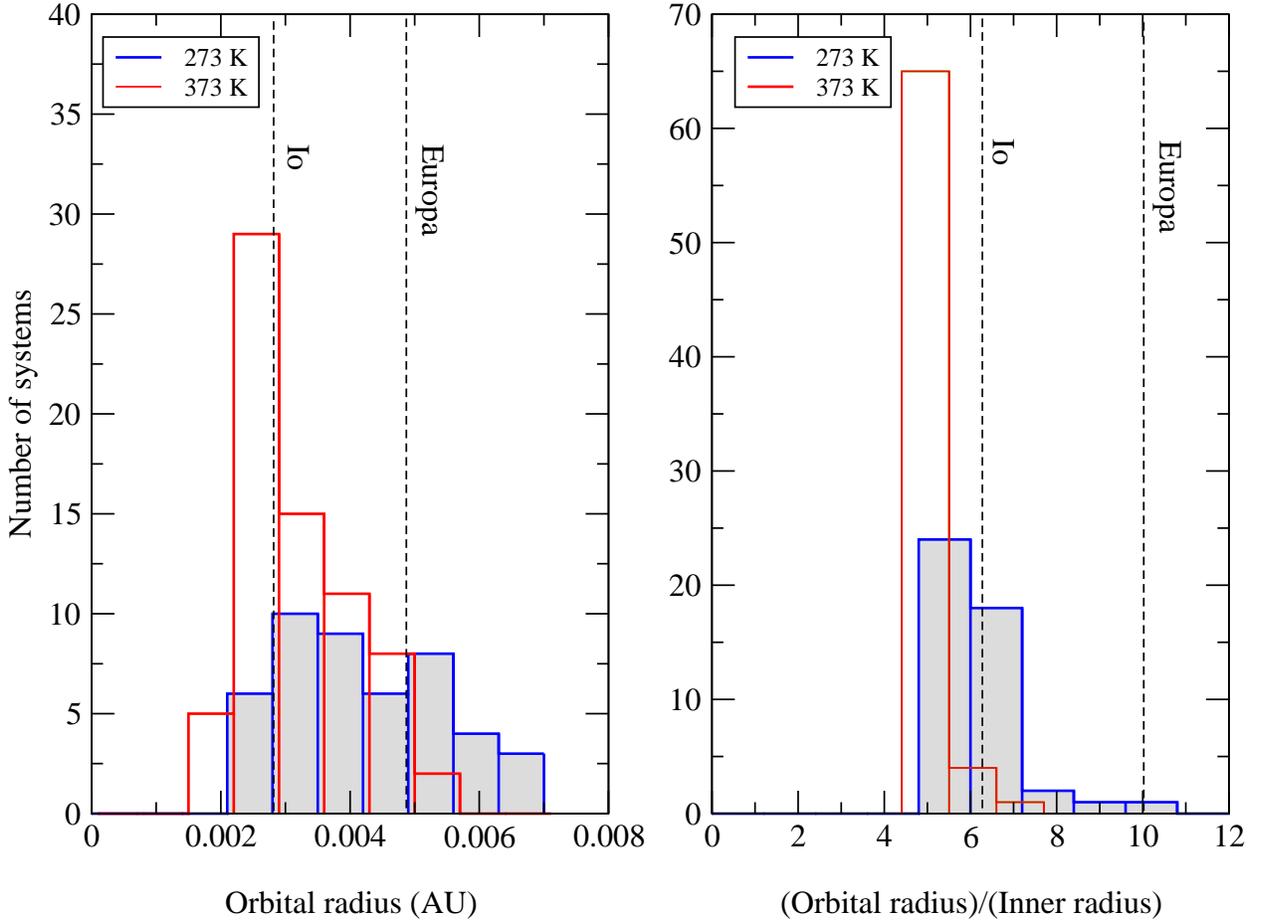}
\caption{Panel (a): the distribution of required moon orbits is plotted for a 0.1M$_{\oplus}$ moon with $e=0.01$ and an assumed $\epsilon=0.62$, such that tidal boosting results in a time-average surface temperature of 273 K (blue outline, grey fill) or 373 K (red outline). Vertical dashed lines indicate the orbital radii of Io and Europa for comparison. Panel (b): the same data is plotted in terms of the ratio of the required  orbital radius to the inner allowed radius ($a_s^{inner}$). 
}
\end{figure}

\begin{figure}
\plotone{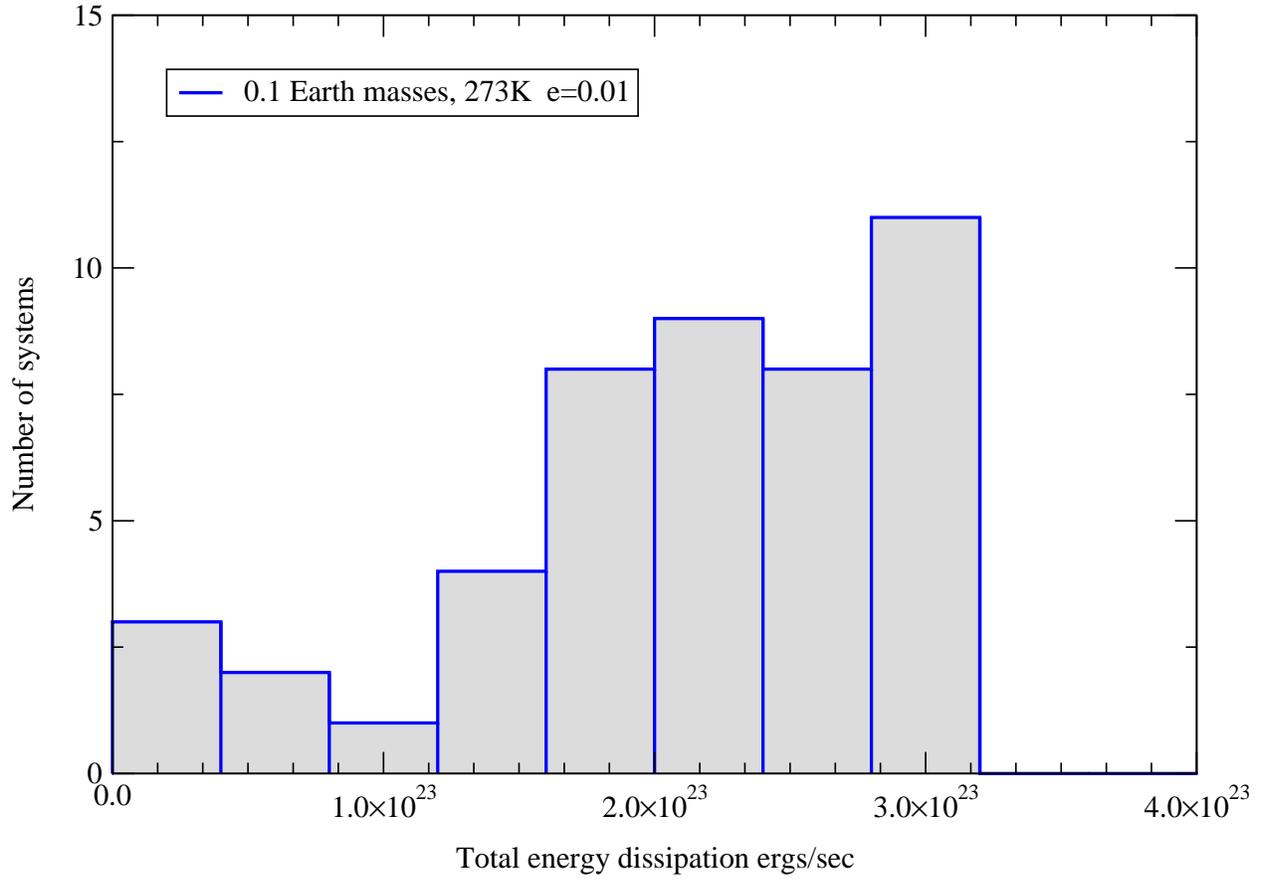}
\caption{Distribution of total energy dissipation for the hypothetical tidally boosted moons with time-averaged temperature of 273 K shown in Figure 7 for the current exoplanet sample. 
}
\end{figure}

\end{document}